\documentclass[twocolumn,graphics,floatfix,a4paper,prl]{revtex4-1}
\usepackage{color}
\usepackage[utf8]{inputenc}
\usepackage{graphicx}
\usepackage{amsmath}
\usepackage{amssymb}
\usepackage{pdfpages}
\makeatletter
\AtBeginDocument{\let\LS@rot\@undefined}
\makeatother

\usepackage{hyperref}
\hypersetup{
  pdfnewwindow=true, colorlinks=true,
  linkcolor=blue, anchorcolor=blue,
  citecolor=blue, filecolor=blue,
  menucolor=blue, urlcolor=blue}

\newcommand{\comment}[1]{\textit{}}
\newcommand{\bit}{\begin{itemize} \setlength{\itemsep}{0ex} \setlength{\topsep}{0ex} } 
\newcommand{\eit}{\end{itemize}}
\newcommand{\be}{\begin{equation}}
\newcommand{\ee}{\end{equation}}
\newcommand{\bea}{\begin{eqnarray}}
\newcommand{\eea}{\end{eqnarray}}
\newcommand{\ba}{\begin{align}}
\newcommand{\ea}{\end{align}}
\newcommand{\SKIP}[1]{}

\newcommand{\ket}[1]{\left| #1 \right\rangle}

\providecommand{\LS}{\ensuremath{{\bf L}\cdot{\bf S}}}
\providecommand{\kp}{\ensuremath{{\bf k}\cdot{\bf p}}}

\providecommand{\bc}{\ensuremath{B_\mathrm{crit}}}

\newcommand{\mb}{\ensuremath{\mathbf}}

\begin{document}
\title{Orbital contributions to the electron $g$-factor in
  semiconductor nanowires}
\author{Georg W. Winkler$^1$} \email{winklerg@ethz.ch}
\author{D\'aniel Varjas$^{2}$} \author{Rafal Skolasinski$^{2}$}
\author{Alexey A. Soluyanov$^{1,3}$} \author{Matthias Troyer$^{1,4}$}
\author{Michael Wimmer$^{2}$} \affiliation{$^1$Theoretical Physics and
  Station Q Zurich, ETH Zurich, 8093 Zurich, Switzerland}
\affiliation{$^2$QuTech and Kavli Institute of Nanoscience, Delft
  University of Technology, 2600 GA Delft, The Netherlands}
\affiliation{$^3$Department of Physics, St.~Petersburg State
  University, St.~Petersburg, 199034 Russia} \affiliation{$^4$Quantum
  Architectures and Computation Group, Microsoft Research, Redmond WA}
\date{\today}

\begin{abstract}
  Recent experiments on Majorana fermions in semiconductor nanowires
  \href{https://dx.doi.org/doi:10.1038/nature17162}{[Albrecht {\em et
      al.}, Nat. 531, 206 (2016)]} revealed a surprisingly large
  electronic Land\'e $g$-factor, several times larger than the bulk
  value --- contrary to the expectation that confinement reduces the
  $g$-factor. Here we assess the role of orbital contributions to the
  electron $g$-factor in nanowires and quantum dots. We show that an
  $\LS$ coupling in higher subbands leads to an enhancement of the
  $g$-factor of an order of magnitude or more for small effective mass
  semiconductors.  We validate our theoretical finding with
  simulations of InAs and InSb, showing that the effect persists even
  if cylindrical symmetry is broken. A huge anisotropy of the enhanced
  $g$-factors under magnetic field rotation allows for a
  straightforward experimental test of this theory.
\end{abstract}

\maketitle
Early electron spin resonance experiments in the 2D electron gas
(2DEG) formed in AlGaAs/GaAs heterostructures found a reduced Land\'e
$g$-factor of electrons~\cite{Stein1983}, which was later
theoretically explained to arise due to the electronic
confinement~\cite{Lommer1985,Ivchenko1992,Ivchenko1998}. It is by now
well established that confinement in a nanostructure leads to a
reduction in the $g$-factor~\cite{Winkler2003,exchange_note} -- the
subband confinement increases the energy gap which is inversely
proportional to $g^* - g_0$, where $g^*$ is the effective and $g_0$
the free electron $g$-factor~\cite{Roth1959,Winkler2003}.
Surprisingly, experiments in InAs \cite{Csonka2008,Petta} and InSb
\cite{Nilson2009,Leo2913} nanowires found $g$-factors surpassing the
corresponding bulk $g$-factors by up to 40\%.

Recently, this discrepancy has attracted interest due to the
experimental discovery of a zero bias conductance peak in
semiconductor nanowires proximity coupled to an $s$-wave
superconductor~\cite{das_majorana_2012,xu_majorana_2012,mourik_majorana_2012,CM_Majorana2016,Zhang2016},
which is believed to be a signature of the Majorana bound
state~\cite{kitaev2001,lutchyn_majorana_2010,oreg_helical_2010} having
possible applications in topological quantum
computation~\cite{top_comp1,top_comp2}. The electron $g$-factor of the
semiconductor nanowire determines the strength of magnetic field
required to trigger the topological phase transition in these
systems. It is desirable to keep the magnetic field low since it also
suppresses superconductivity, and thus a large $g$-factor
semiconductor is desired. Furthermore, Majorana proposals based on
magnetic textures~\cite{magnetic_textures0, magnetic_textures1,
  magnetic_textures2} and various spintronic
devices~\cite{spintronics_review} require large $g$-factors. Small
band-gap semiconductors like InAs and InSb are therefore the materials
of choice for Majorana nanowires, having large $g$-factors and strong
spin-orbit coupling (SOC).

In a recent experiment with InAs nanowires $g$-factors~\footnote{We
  measure the $g$-factors in units of the Bohr-magneton
  $\mu_B = \frac{e\hbar}{2m_0}$ and use the sign convention where the
  free electron $g$-factor is $g_0 \approx +2$.}  more than three
times larger than the bulk $g$-factor
(${g^*_\mathrm{InAs}=-14.9}$~\cite{Pidgeon1967,Winkler2003}) were
measured~\cite{CM_Majorana2016}.  Moreover, it was found that the
$g$-factor depends very strongly on the chemical potential $\mu$ tuned
by the gate potential~\cite{CM_private_communication}. For low $\mu$
small $g$-factors where found which can be explained by the bulk
$g$-factor of InAs. The anomalously large $g$-factors have been only
detected at high chemical potential $\mu$.

In this work, we present a mechanism that can lead to very large
$g$-factors in higher subbands of nanowires and similarly shaped
nanostructures. With this we can explain both the large $g$-factors
observed in
Refs.~\onlinecite{Csonka2008,Petta,Nilson2009,CM_Majorana2016}, and
the chemical potential dependence~\cite{CM_private_communication}. In
particular, we find that the orbital angular momentum in the confined
nanostructure plays a crucial role. The lowest conduction
subband/state is characterized by no or only small orbital angular
momentum. In this case the usual reasoning applies and confinement
does lead to a reduction of the $g$-factor. Higher subbands/states,
however, can have nonzero orbital angular momentum in an approximately
cylindrical structure. Due to strong SOC in small band-gap
semiconductors one finds an $\LS$-type spin alignment if the orbital
angular momentum $\mathbf{L}$ is nonzero. Kramers pairs of opposite
orbital angular momentum form at $B=0$, and thus the $g$-factor
obtains an additional contribution resulting from the coupling of the
orbital angular momentum to the magnetic field.  A similar orbital
enhancement of the g-factor is known from the theory of the hydrogen
atom~\cite{landau1981quantum} and has also been observed in carbon
nanotubes~\cite{Kuemmeth2008, Laird2015}. However, due to the small
effective mass the $g$-factor enhancement can be orders of magnitude
larger in the semiconducting structures investigated here.

\paragraph{Cylindrical symmetry} --- We start by considering
cylindrical nanowires and estimate the maximally achievable $g$-factor
for subbands as a function of their orbital angular
momentum. Initially, we assume independent SU(2) spin rotation
symmetry (no SOC) and time-reversal (TR) invariance without magnetic
field. We then introduce magnetic field parallel to the wire, thus
preserving the rotational invariance (both in real space and spin)
around the axis of the wire ($z$ direction in the following).

As the wire is translationally invariant in the $z$ direction, and the
conduction band minimum is at $k_z = 0$, we restrict to $k_z = 0$ in
the following and investigate the wavefunction in the $xy$ plane
only. As a consequence of separate real space and spin rotation
symmetries, the states can be classified by their orbital angular
momentum $L_z = 0, \pm \hbar, \pm 2\hbar,$ etc. and spin
$S_z = \pm \frac{\hbar}{2}$ (for brevity we drop the $z$ subscript in
the following and use the lower case letters for angular momentum in
units of $\hbar$). The lowest subband is twofold spin degenerate
$\ket{l=0,s=\pm\frac{1}{2}}$, higher subbands with $l \ne 0$ being
fourfold $\ket{\pm |l|,\pm\frac{1}{2}}$.

In a simple quadratic band with an effective mass $m^*$, the momentum
and electrical current are related as
${\mb J} = \frac{e}{m^*} {\mb p}$.  Using the orbital angular momentum
$\mathbf{L} = {\mb r}\times{\mb p}$ the orbital magnetic moment is
expressible as
\begin{equation}
  {\mb M}_\mathrm{o} = \frac{1}{2} {\mb r}\times {\mb J} = -\frac{e}{2
    m^*} {\mb L} =  -\frac{m_0}{m^*} \mu_B l {\mb e}_z.
  \label{eq:M_o}
\end{equation}
We see that the orbital magnetic moment is enhanced by the low
effective mass of the bands. Because of the fourfold degeneracy, we
cannot unambiguously calculate $g$-factors and thus next include
spin-orbit coupling.

\begin{figure}
  \includegraphics[width = \linewidth]{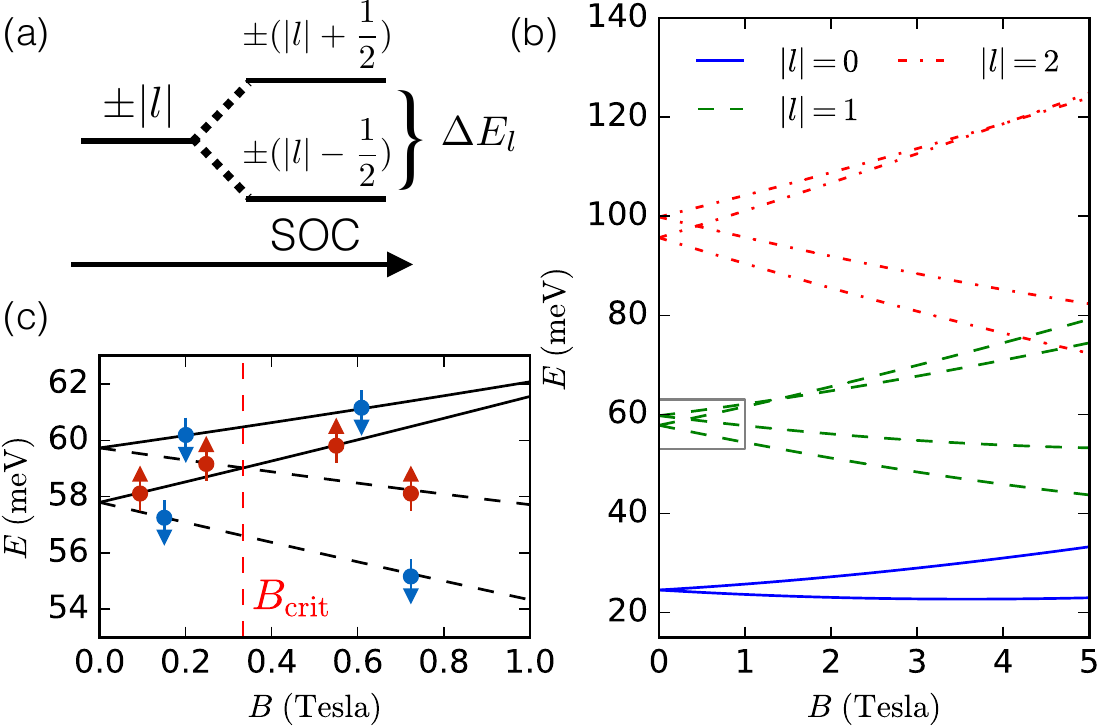}
  \caption{\label{fig:splitting}(a) Evolution of the energy levels at
    $k_z = 0$ in cylindrical symmetry when SOC is turned on. (b)
    Energy levels of a cylindrical InSb wire with 40 nm diameter in an
    axial magnetic field. (c) Zoom in on the $|l|=1$ states marked by
    the gray rectangle in (b). Dashed lines are $l=+1$, and solid
    lines $l=-1$, states. The spin alignments are marked by the small
    arrows and the vertical dashed red line marks $\bc$.}
\end{figure}

With SOC the orbital and spin angular momentum is no longer separately
conserved, but the total angular momentum $f_z = l_z + s_z$ is still
conserved and takes half-integer values. Without magnetic field the
system is TR invariant. As angular momentum is odd under TR, the
degenerate Kramers-pairs have opposite $f$. Turning on SOC splits the
fourfold degeneracy of the $l \ne 0$ subbands into two degenerate
pairs: $\ket{+|l|,+\frac{1}{2}}$ and $\ket{-|l|,-\frac{1}{2}}$ stay
degenerate ($f = \pm (|l|+\frac{1}{2})$) and so do
$\ket{+ |l|,-\frac{1}{2}}$ and $\ket{-|l|,+\frac{1}{2}}$
($f = \pm (|l|-\frac{1}{2})$), as shown in
Fig.~\ref{fig:splitting}~(a). Even though the orbital and local
angular momenta are no longer separately conserved their expectation
values remain similar for realistic SOC strengths.

The magnetic field ${\mb B}$ couples to the total magnetic moment
${\mb M} = {\mb M_{o}} - g^* \frac{e}{2m_0} {\mb
  S}$~\cite{Ivchenko1998}. Using Eq.~\eqref{eq:M_o}, the Zeeman
splitting of a Kramer's pair $\ket{\pm |l|, +\frac{1}{2}}$ and
$\ket{\mp |l|,-\frac{1}{2}}$ for a magnetic field in $z$-direction is
given by
$\Delta E_\mathrm{Zeeman} = \mu_B (g^* \pm 2 \frac{m_0}{m^*} |l|)
\frac{B_z}{2}$ and the resulting effective $g$-factor can be read off
\begin{equation}
  g_{|l|\pm\frac{1}{2}} = g^* \pm 2 \frac{m_0}{m^*} |l|.
  \label{eq:gfactor}
\end{equation}
Below we will see from numerical simulation that this is a good
approximation even in a less ideal case.

This result is analogous to the well known Land{\'e} $g$-factor of the
Hydrogen atom when taking relativistic SOC into account: the splitting
induced by weak external magnetic field has contributions from both
the orbital and spin angular momentum~\cite{landau1981quantum}. This
effect is amplified in semiconductor nanostructures because the small
effective mass increases both the orbital magnetic moment and the bulk
$g$-factor $g^*$.

\paragraph{Wire simulations} --- We next validate our theoretical
findings with simulations of nanowires using an eight-band $\kp$-model
for zincblende
semiconductors~\cite{Kane1957,Foreman1997,Winkler2003}. At first, we
assume perfect cylindrical symmetry of a nanowire, grown in 001
direction, and employ the axial
approximation~\cite{SercelPRB1990,SercelPRL1990,Cakmak2003,supplementary}. In
this case, the wavefunctions can be written as~\cite{Galeriu2005}
\begin{equation}
  \psi(\rho , \phi, z) = \sum_n g_n(\rho,z) e^{il_n\phi} |u_n \rangle,
  \label{eq:separation}
\end{equation}
where $|u_n \rangle$ are the basis states of the 8-band $\kp$
Hamiltonian with local angular momentum $j_n$ \footnote{Here $j$ takes
  the role of $s$ in the previous argument, as in these materials the
  $p$-type orbitals have nonzero local orbital angular momentum and
  are treated as spin-$3/2$ degrees of freedom.}. Since the
Hamiltonian conserves the total angular momentum $f$ one obtains the
orbital part of each component as $l_n = f - j_n$. If we furthermore
focus on an infinite wire in the $z$-direction the problem is reduced
to a 1D boundary value problem in $\rho$ which we solve using the
finite difference method~\cite{supplementary}.

\begin{figure}
  \includegraphics[width =
  \linewidth]{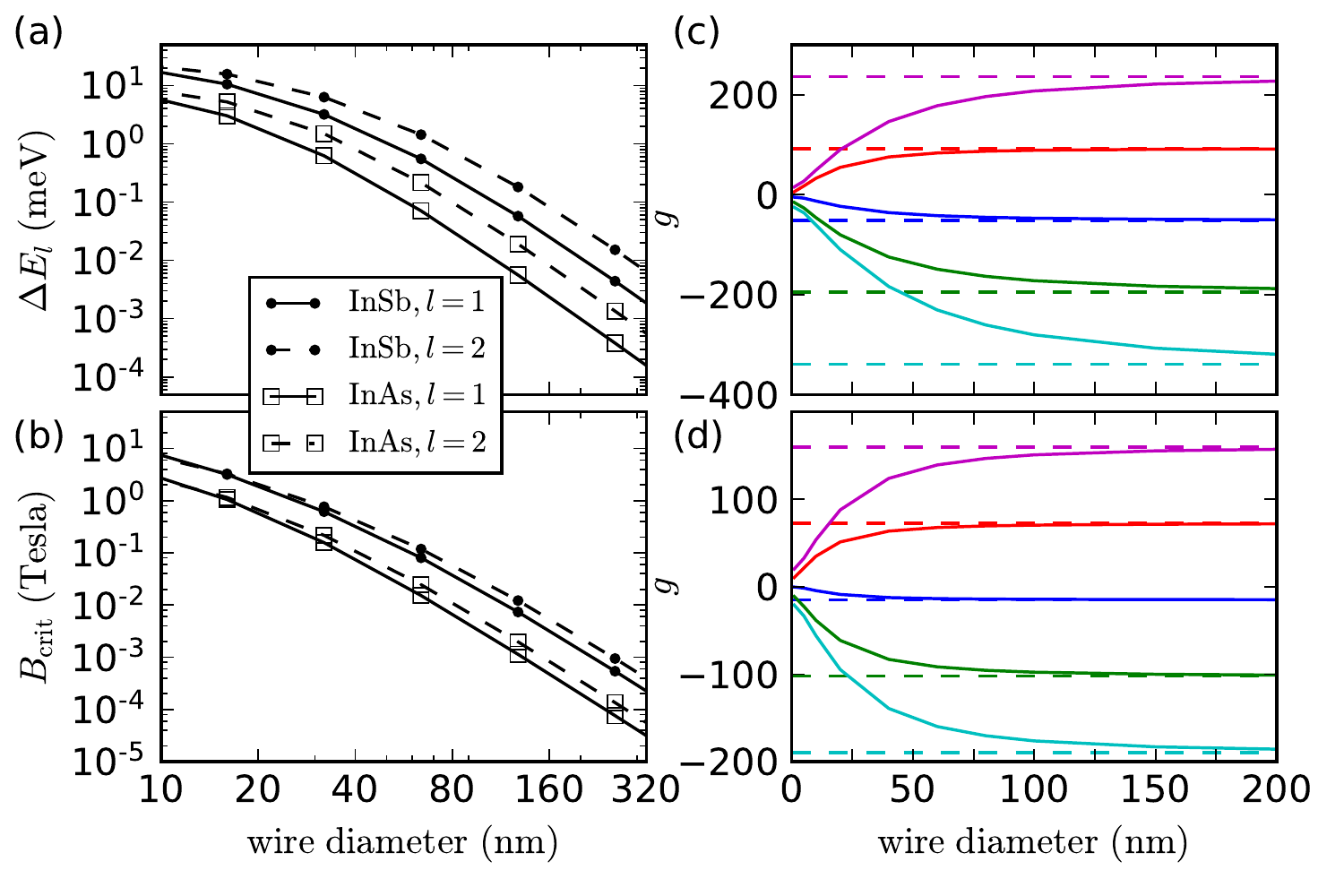}
  \caption{\label{fig:diameter}(a) Diameter dependence of the SOC
    splitting $\Delta E_1$ and $\Delta E_2$ for InSb and InAs
    wires. (b) Diameter dependence of the critical magnetic field
    $\bc$ defined in Fig.~\ref{fig:splitting}~(b). (c)((d)) Effective
    $g$-factors at infinitesimal magnetic field of the first five
    subbands of an InSb (InAs) wire: $l=0,|f|=\frac{1}{2}$ (blue),
    $|l|=1,|f|=\frac{1}{2}$ (green), $|l|=1,|f|=\frac{3}{2}$ (red),
    $|l|=2,|f|=\frac{3}{2}$ (cyan) and $|l|=2,|f|=\frac{5}{2}$
    (magenta). The dashed lines in the corresponding colors are the
    prediction of Eq.~\eqref{eq:gfactor} where we substituted bulk
    values.}
\end{figure}

Figure~\ref{fig:splitting} (b) shows the subband edges of an InSb
nanowire of 40 nm diameter. At $B = 0$ one generically finds the
lowest conduction subband to originate from the $|l|=0$ state without
SOC. At higher energy there are the $|l|=1$ and $|l|=2$ states and
then another $|l|=0$ state with a higher radial quantum number (not
shown). This order of states is generic as long as the conduction band
is approximately quadratic~\cite{Robinett}. Figure~\ref{fig:splitting}
(c) zooms in on the $|l|=1$ subbands. Due to SOC the $|f|=\frac{3}{2}$
and $|f|=\frac{1}{2}$ states are split at $B=0$ by
$\Delta E_1 \approx 2\,\mathrm{meV}$. If a magnetic field $B < \bc$
(see Fig.~\ref{fig:splitting}~(c)) is turned on a splitting between
states of opposite orbital angular momentum $l$ is observed and thus,
enhanced $g$-factors according to Eq.~\eqref{eq:gfactor}. However,
when the magnetic field is large, $B > \bc$, states of the same
orbital angular momentum bundle together and their relative slope with
respect to $B$ corresponds to the normal $g$-factor without orbital
contributions. Thus a splitting $\Delta E_l$ is a crucial ingredient
for enhanced $g$-factors.

Figure~\ref{fig:diameter} shows the dependence on the diameter of the
nanowire.  From the $\Delta E_l$ dependence it is evident that the
wire cannot be made too thick to experimentally observe the effect
with a detectable energy scale, e.g. to distinguish the split energy
levels using Coulomb
oscillations~\cite{Leo_coulomb}. Figures~\ref{fig:diameter}~(c) and
(d) show that at large wire diameters Eq.~\eqref{eq:gfactor} is
reproduced perfectly by numerics, but for small diameters the
$g$-factor enhancement is reduced by the confinement. Thus, the
optimal diameter range where enhancement of the $g$-factor is strong
and at the same time $\Delta E_l$ and $\bc$ are large enough is in
between 10 and 100 nm. We see that the $g$-factors of higher subbands
can be very large --- enhancements of an order of magnitude compared
to the bulk $g$-factor are possible.

The splitting $\Delta E_l$ is generic if SOC is present, since in a
typical semiconductor wire with SOC there is no symmetry that would
protect the degeneracy between states of different total angular
momentum. The conduction band of zincblende semiconductors has a
purely $s$-orbital character at the $\Gamma$-point of the Brillouin
zone (BZ), which is insensitive to SOC. Thus, also the conduction
subbands of a zincblende nanowire are mostly derived from
$s$-orbitals. Any nonzero splitting $\Delta E_l$ results from $p$-like
hole contributions to the conduction band due to confinement. This
explains why the splitting in the conduction band is so small compared
to the split-off energy of the valence bands $\Delta$, which is 0.81
eV for InSb and 0.38 eV for InAs~\cite{Winkler2003}.

\begin{figure}
  \includegraphics[width = \linewidth]{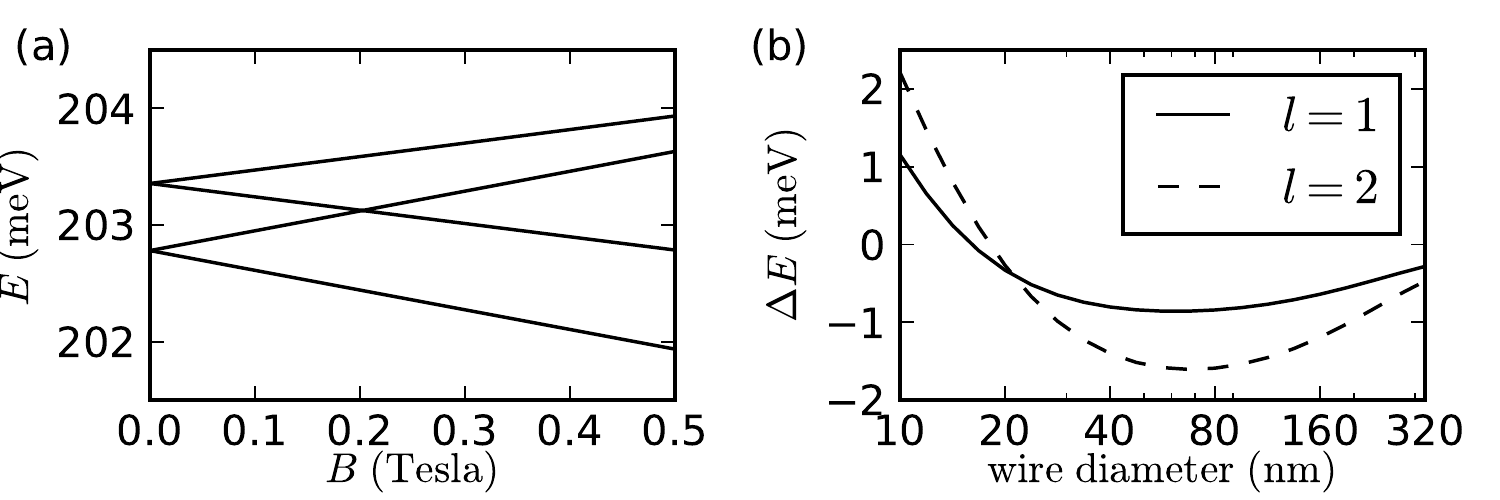}
  \caption{\label{fig:combined} (a) Energy levels of the lowest
    $|l|=1$ states as a function of $B$ in a tight-binding simulation
    of a hexagonal InSb wire of 20.1 nm diameter, grown in the 111
    direction. (b) SOC splitting as a function of diameter in a
    cylindrical wurtzite InAs wire.  }
\end{figure}

Since $\Delta E_l$ results from the scattering of states at the
surface of the wire the boundary conditions impact the numerical
value, and even the sign, of $\Delta E_l$~\cite{supplementary}. Abrupt
boundaries can be problematic in $\kp$ simulations~\cite{Rodina2002},
therefore, we use tight-binding (TB) simulations to check the
robustness of our results. The effective tight-binding Hamiltonian is
generated from the first-principles $s$ and $p$-like Wannier
functions~\cite{mostofi2008wannier90}, calculated using the Vienna
{\it ab initio} simulation package
(VASP)~\cite{kresse_vasp_1993,kresse_vasp_1994,kresse_vasp_1996,kresse_vasp1_1996}
with the projector augmented-wave method~\cite{PAW1,PAW2}, a cut-off
energy of 300~eV, a 8$\times$8$\times$8 Monkhorst-Pack mesh and using
the HSE06 hybrid functional~\cite{kresse_hse06, heyd_hybrid_2003,
  heyd_hybrid_2004}. Furthermore, the TB model includes the
Dresselhaus term which was neglected for the zincblende $\kp$
simulations since its effect is found to be very
small~\cite{supplementary}. In Fig.~\ref{fig:combined}~(a) we show the
magnetic field dependence of the $|l|=1$ subbands in a hexagonal InSb
wire. The $g$-factors of -59 and +40 and
$\bc\approx 0.2\,\mathrm{Tesla}$ agree qualitatively with the
$\kp$-results.

While in zincblende wires boundary effects are dominating, in wurtzite
wires the situation is different: There, the conduction band has a
mixed $s$ and $p$-character. Thus, wurtzite wires have an intrinsic
splitting independent of
confinement~\cite{christensen_wurtzite}. Using a $\kp$-model for
wurtzite semiconductors~\cite{Faria_wurtzite}, we find a nearly
size-independent $\Delta E_l$ of order 1 meV for [0001] grown wurtzite
InAs wires for experimentally used diameters of 40 to 160
nm~\cite{peter_nanowires}, see Fig.~\ref{fig:combined}~(b). At very
large wire diameters $> 200$\,nm the confinement induced subband
splitting becomes smaller than $\Delta E_l$, leading to a reduction of
$\Delta E_l$, and at very small diameters $< 20$\,nm the cubic
Dresselhaus term dominates over the linear Rashba term, causing a sign
change in $\Delta E_l$~\cite{supplementary, Gmitra_wurtzite}.

\begin{figure}
  \includegraphics[width = \linewidth]{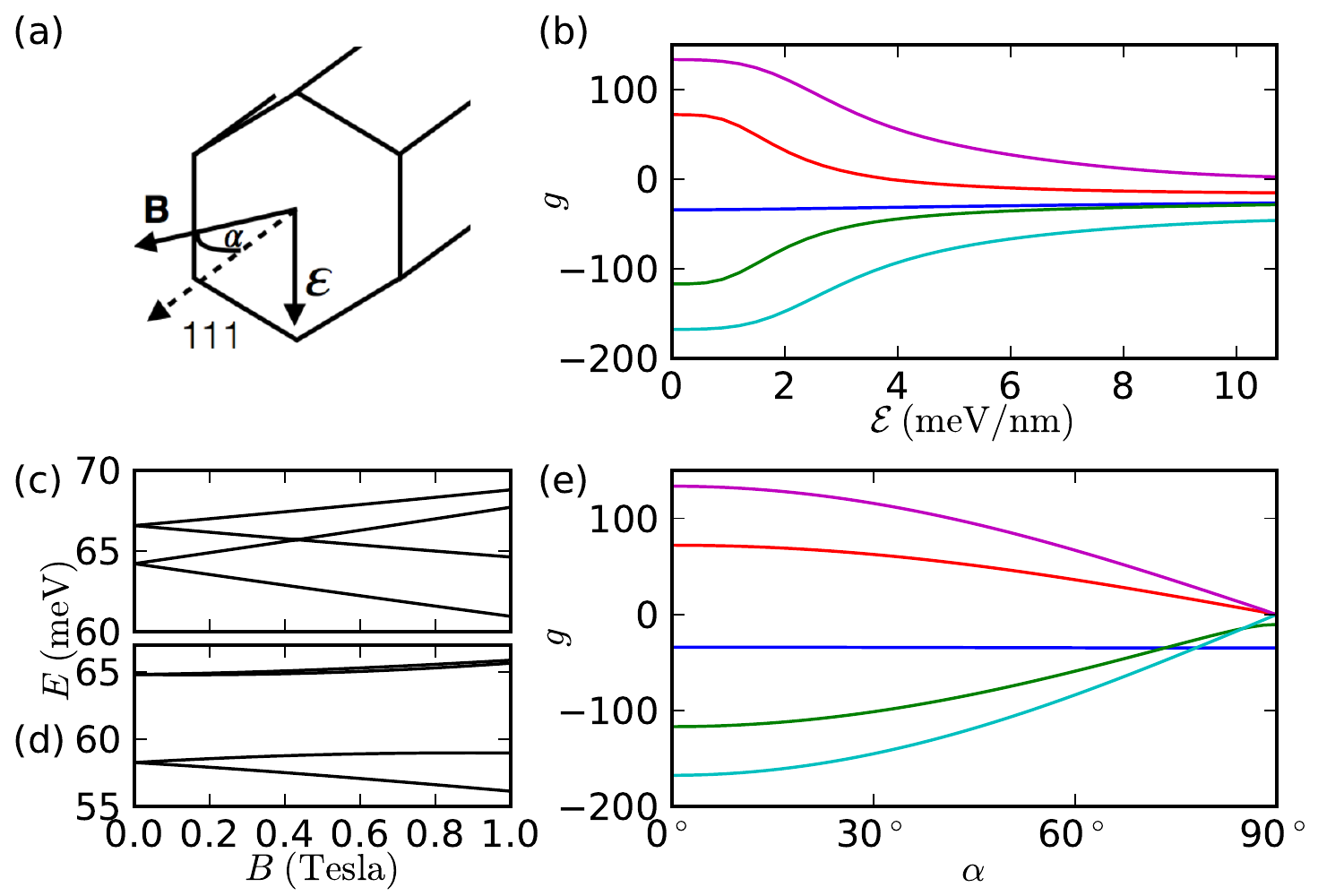}
  \caption{\label{fig:hexagonal} (a) Magnetic and electric field
    directions in the hexagonal 111 wire.  (b) The $g$-factors
    measured at 0.2 Tesla ($\alpha=0$) of a hexagonal InSb wire with
    40 nm diameter as a function of a perpendicular electric field.
    (c) ((d)) Energy levels of the $|l|=1$ states as a function of $B$
    at an electric field of ${\mathcal{E}=0\,\mathrm{meV/nm}}$
    (${\mathcal{E}=3\,\mathrm{meV/nm}}$). (e) The $g$-factors as a
    function of $\alpha$ measured at 0.2 Tesla in a hexagonal InSb
    wire with 40 nm diameter. In (b) and (e) the color code is the
    same as in Fig.~\ref{fig:diameter}~(c/d).}
\end{figure}

\paragraph{Symmetry breaking} --- We now consider the effects of
broken cylindrical symmetry and solve the full 2D cross section of
hexagonal zincblende wires, grown in the 111 direction, using a 2D
discretization of the $\kp$-model~\cite{supplementary,
  orbital_majorana}. We allow for symmetry breaking by electric field
and off-axis magnetic field, see Fig.~\ref{fig:hexagonal}~(a) for the
definitions of the relevant directions.
In experimental situations, the symmetry is generally broken by
electric fields, e.g. due to the backgate for tuning the electron
density in the
wire~\cite{Csonka2008,Nilson2009,CM_Majorana2016,CM_private_communication}.
We find that, especially in higher subbands, the enhanced $g$-factors
are quite robust to an external electric field.

In Fig.~\ref{fig:hexagonal} (b-d) we simulate a hexagonal InSb wire,
of 40 nm diameter, in a perpendicular external electric field
$\mathcal{E}$. The point group of the wire at $\mathcal{E}=0$ is
$C_{3v}$ and crossings between states of different angular momentum
are protected, as illustrated in Figure~\ref{fig:hexagonal}~(c). At
nonzero $\mathcal{E}$ the different angular momentum eigenstates
hybridize, which reduces their orbital angular momentum expectation
value. However, as shown in Figs.~\ref{fig:hexagonal}~(b) and (d), the
orbital contribution to the $g$-factor remains very significant until
very large fields are applied. Bands with larger values of $|l|$ have
larger splitting $\Delta E_l$ and, therefore, the orbital contribution
to their $g$-factors is more robust and can remain significantly
larger than the bulk $g$-factor until large electric fields, e.g. see
the cyan and magenta lines corresponding to $|l|=2$ in
Fig.~\ref{fig:hexagonal}~(b).

The electron $g$-factor anisotropy in the magnetic field of 2DEGs is
well
established~\cite{Ivchenko1992,Kesteren1990,Peyla1993,Winkler2003}. In
our case of orbitally enhanced $g$-factors in nanowires we expect an
even stronger anisotropy. Indeed, the electron spins in subbands with
$l\ne 0$ feel a very strong orbital magnetic field that aligns them
(anti-) parallel to the wire axis. Therefore, a perpendicular magnetic
field first needs to overcome this orbital effect to create a Zeeman
splitting of the states~\cite{Kuemmeth2008, Laird2015}.

This is illustrated in Fig.~\ref{fig:hexagonal}~(e), where we simulate
a hexagonal InSb wire of 40 nm diameter in a magnetic field of 0.2
Tesla. We show there the g-factor as a function of the angle $\alpha$
between the magnetic field and the nanowire axis. While the $g$-factor
of the lowest $l=0$ subband is unaffected by the direction of
$\mathbf{B}$, the $g$-factor for bands with $l \ne 0$ almost vanishes
for perpendicular magnetic field. This strong anisotropy of the
electron $g$-factor can be used in experiments to prove the important
role of orbital angular momentum in nanowires.

\begin{figure}[t]
  \includegraphics[width = \linewidth]{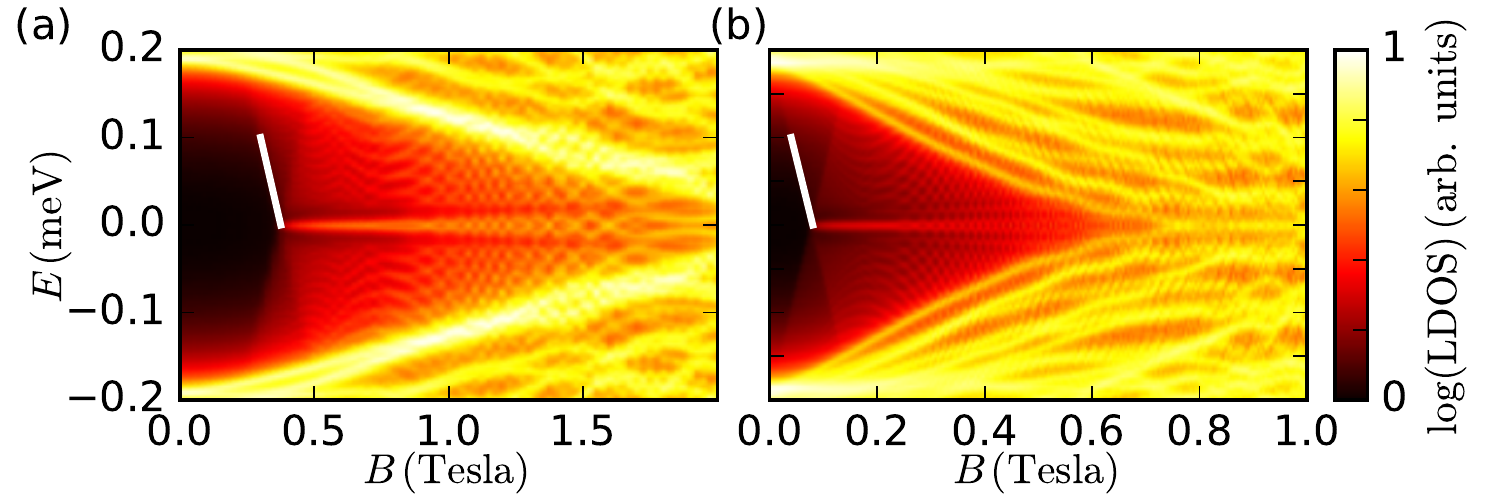}
  \caption{\label{fig:majorana} (a) ((b)) show the local density of
    states (LDOS) at the end of an InAs wire with 40 nm diameter and
    2172 nm length in an electric field of
    ${\mathcal{E}=1.2\,\mathrm{meV/nm}}$ and proximity effect induced
    superconducting pairing $\Delta=0.2\,\mathrm{meV}$. The chemical
    potential $\mu = 39.6\,\mathrm{meV}$ ($\mu = 68.5\,\mathrm{meV}$)
    is tuned to the $|l|=1$ ($|l|=2$) subbands. The slope of the
    whites lines amount to a $g$-factor of 23 (43). }
\end{figure}

In a Majorana wire circular symmetry breaking by gate potentials and
band bending is mandatory to create a Rashba effect in the wire
\cite{lutchyn_majorana_2010, oreg_helical_2010, CM2015}. The results
shown above suggest that even in such an environment orbital effects
still dominate the $g$-factors of certain subbands in wires.  This is
illustrated in Fig.~\ref{fig:majorana} (a) and (b), where we simulate
an InAs wire proximity coupled to an Al superconductor (see the
Supplemental Material~\cite{supplementary} for the details of the
simulation).  When the chemical potential is tuned to the $|l|=1$ and
$|l|=2$ subbands, the $g$-factors, extracted from the slope of the
Majorana state forming Andreev bound state, are 23 and
43~\footnote{Our simulations do not include the renormalization
  effects of the
  superconductor~\cite{renormalization,renormalization1}, which could
  lead to a reduction of the resulting $g$-factor.},
respectively. These $g$-factors are significantly larger than the bulk
$g$-factor of InAs, thus reproducing the experimental result of
Ref.~\onlinecite{CM_Majorana2016}.

\paragraph{Conclusions and Outlook --- }
In summary, we have provided a theory for the previously unexplained
large $g$-factors observed in nanowires. Our findings help to better
understand and optimize Majorana experiments.  Similar results apply
to quantum dots. For cylindrical quantum dots we find that orbital
$g$-factor enhancements are still significant if the length of the dot
is much shorter than its diameter, see the Supplemental
Material~\onlinecite{supplementary} for more details. Due to the
observed robustness of the effect, it also applies in irregularly
shaped quantum dots and can explain $g$-factor fluctuations there.

\acknowledgments
{\it Acknowledgments.} We would like to thank L Kouwenhoven, S
Vaitiek{\.e}nas, MT Deng, CM Marcus, K Ennslin, TD Stanescu, AE
Antipov, E Rossi, and RM Lutchyn for useful discussions and QS Wu for
providing first-principles derived tight-binding models. This work was
supported by Microsoft Research, the Netherlands Organization for
Scientific Research (NWO), the Foundation for Fundamental Research on
Matter (FOM), the European Research Council through ERC Advanced Grant
SIMCOFE, the Swiss National Science Foundation and through the
National Competence Centers in Research MARVEL and QSIT.

\bibliography{literature}
\newpage
\mbox{}
\newpage
\includepdf[pages=1]{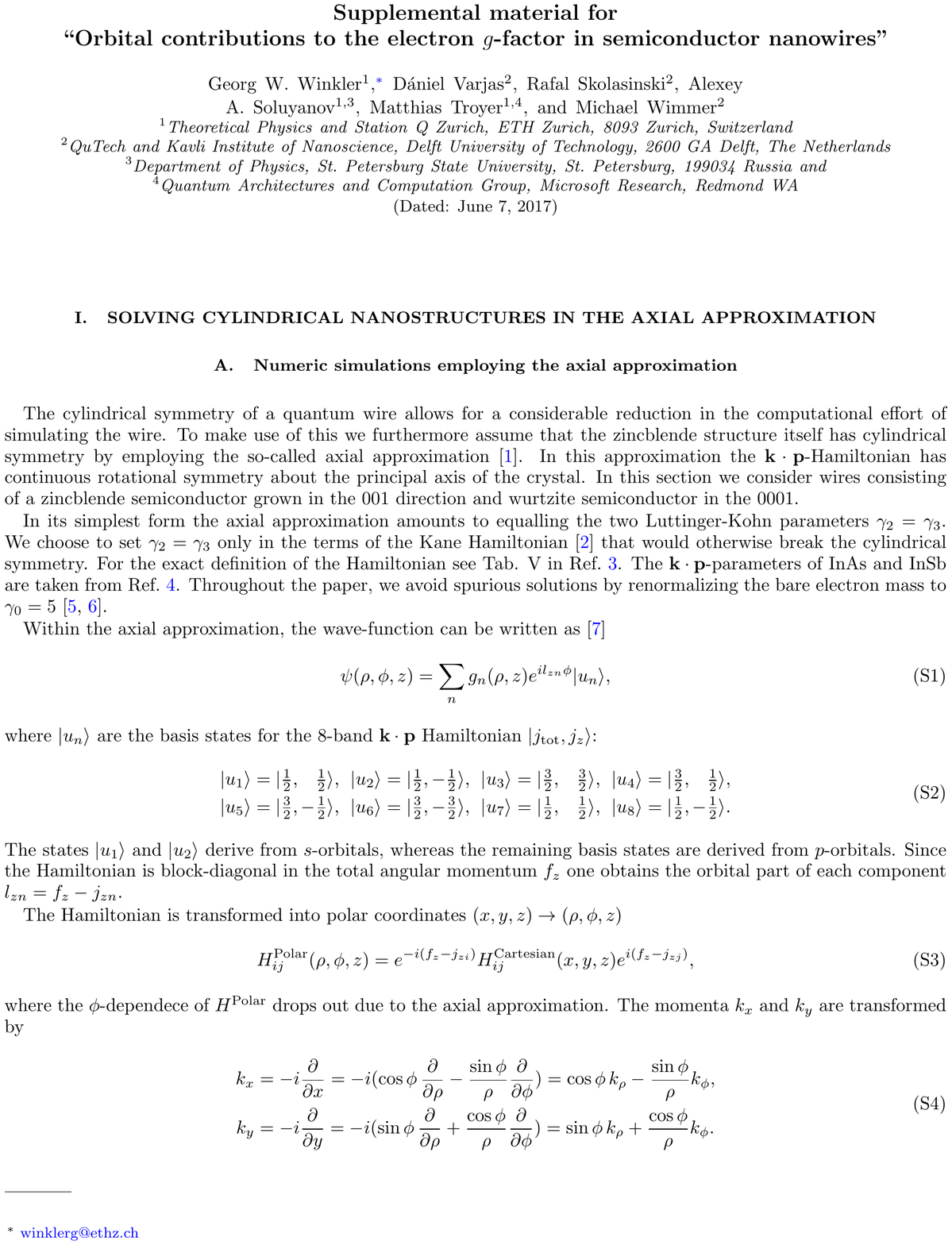}
\mbox{}
\newpage
\includepdf[pages=2]{supp_mat}
\mbox{}
\newpage
\includepdf[pages=3]{supp_mat}
\mbox{}
\newpage
\includepdf[pages=4]{supp_mat}
\mbox{}
\newpage
\includepdf[pages=5]{supp_mat}
\mbox{}
\newpage
\includepdf[pages=6]{supp_mat}
\mbox{}
\newpage
\includepdf[pages=7]{supp_mat}
\mbox{}
\newpage
\includepdf[pages=8]{supp_mat}
\mbox{}
\newpage
\includepdf[pages=9]{supp_mat}
\mbox{}
\newpage
\includepdf[pages=10]{supp_mat}
\mbox{}
\newpage
\includepdf[pages=11]{supp_mat}
\mbox{}
\newpage
\includepdf[pages=12]{supp_mat}

\end{document}